\documentclass{article}

\usepackage[english]{babel}

\usepackage[letterpaper,top=2cm,bottom=2cm,left=3cm,right=3cm,marginparwidth=1.75cm]{geometry}

\usepackage{amsmath}
\usepackage{graphicx}
\usepackage{tabularx}
\usepackage[numbers]{natbib}
\usepackage[colorlinks=true, allcolors=blue]{hyperref}
\usepackage{amssymb}
\usepackage{float}
\usepackage{setspace}
\usepackage{multirow}
\usepackage{bbm}
\usepackage[skip=10pt plus1pt]{parskip}

\usepackage{authblk}
\title{Comparison of probabilistic nowcasts and forecasts of SARS-CoV-2 variant proportions made by hierarchical multinomial linear regression models }
\author[1]{Isaac MacArthur}
\author[1]{Thomas Robacker}
\author[2]{Evan L. Ray}
\author[2]{Benjamin W. Rogers}
\author[2]{Nicholas G. Reich}
\author[1]{Maryclare Griffin}
\affil[1]{Department of Mathematics and Statistics, University of Massachusetts, Amherst}
\affil[2]{School of Public Health and Health Sciences, University of Massachusetts, Amherst}
\begin{document}
\maketitle

% Variant or clade?
% Target date vs. nowcast date
    %% submission date: date on which a model is fit and a prediction/nowcast is made
    %% target date: date for which a nowcast or forecast is made
% rewrite key contribution to be less staty

% Target Journals
% - Plos Computational Biology?

% Note: Acknowledge Becky and other variant hub folks

%\tableofcontents

\clearpage 
{\bf Key Contribution}: 
We build off of work exploring the ability of models to capture the spread of different strains of infectious diseases. In particular, we define a class of popular statistical regression models and rigorously test their ability to accurately predict SARS-CoV-2 (COVID-19) variant proportions at the US state level. We evaluate the models both in terms of their average prediction error and their prediction uncertainty.  
\begin{abstract}
Nowcasting and forecasting of infectious diseases have become increasingly important since the SARS-CoV-2 pandemic. In particular, methods for modeling the composition of circulating variants at a given time have seen more use in part due to a large increase in the frequency of genomic sequencing conducted as a part of routine surveillance. However, methods must take into account that locations have different amounts of data and sometimes have different trends. We discuss hierarchical multinomial logistic regression (HMLR), a commonly used method for forecasting SARS-CoV-2 variants, which allows for data sharing across locations. We show how it has been used in the literature, and define a class of HMLR models for SARS-CoV-2 variant nowcasting and forecasting. We rigorously test a subset of this class of models using the framework of the US SARS-CoV-2 Variant Nowcast Hub, a collaborative modeling project that launched in 2024. We created two years of weekly predictions based on retrospective datasets, with the prediction dates ranging from Wednesday, August 3, 2022, to Wednesday, August 7, 2024. We tested 12 HMLR models against a baseline model on these datasets. We found that the HMLR models outperformed the baseline both in terms of probabilistic accuracy, as measured by the energy score, as well as point accuracy, as measured by the Brier score. Overall, we find that HMLR models perform best with respect to the baseline model in locations with more data, and more complex HMLR models also showed more improvement in those high-data locations; however, there was no one best model across all metrics, and simpler HMLR models perform better in low-data locations. We find that HMLR models perform well in practice for nowcasting and forecasting SARS-CoV-2 variants.      
\end{abstract}
\section{Introduction}
Forecasting and nowcasting of pathogen prevalence, as measured through genomic sequence data, has taken on increased importance due to the rise of new epidemic threats such as SARS-CoV-2 (the virus that causes COVID-19). Pathogens evolve and change over time, leading to the emergence of different variants, subtypes, or clades. As was seen during the COVID-19 pandemic, different variants can cause different public health burdens \citep{hamed2021global, young2021association}. Therefore, accurate assessments of variant proportions in specific locations are one component of a well-informed public health response to outbreaks.
This paper describes a class of hierarchical multinomial regression models that have been designed to nowcast (predict incomplete observations in the present and recent past) and forecast (predict observations in the future) variant proportions across multiple locations and compares the performance of different models in this class. 
This hierarchical structure is useful when there are wide disparities in quantities of data across locations or when not all locations observe all of the possible targets we are interested in.  
A key feature of these models is that the hierarchical structure allows for adaptive data sharing across locations, which lets locations with more data inform trends in locations with less data while still allowing for variation across locations \citep{gelman2006multilevel, woltman2012introduction}. Additionally, these models produce probability distributions for all locations that reflect nowcast and forecast uncertainty.

\begin{figure}[h]
\begin{center}
\includegraphics[width=\columnwidth]{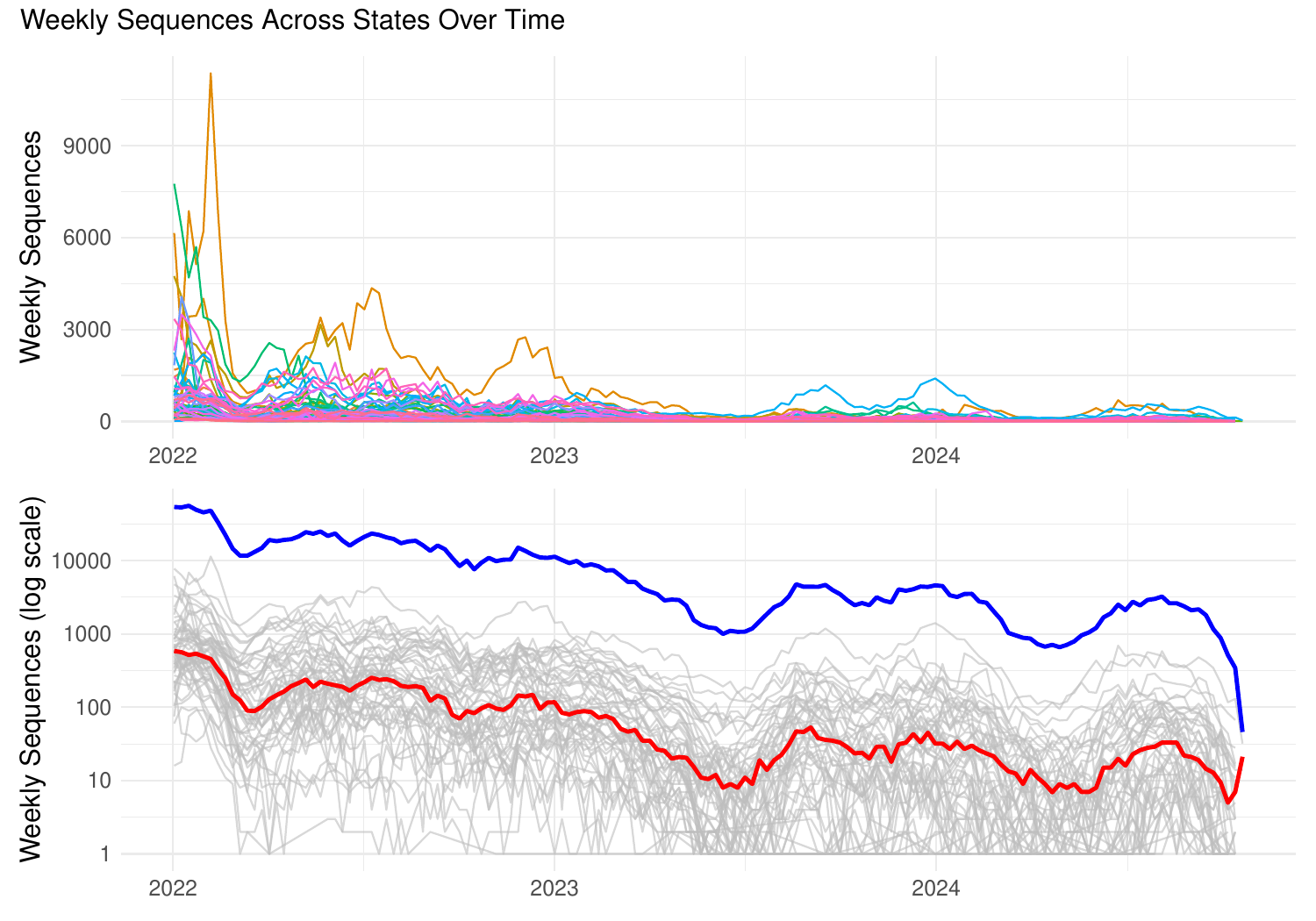}
\caption{Number of sequences per week for each of the 50 states, Puerto Rico, and Washington D.C. from the first MMWR week of 2022 to the week of November 4, 2024. {\bf Top}: Data shown on natural scale; each line is a state. There is a large drop in the number of sequences after 2022 and the magnitudes of sequence counts vary widely across states. {\bf Bottom}: Data shown on the logarithmic scale. Each state is shown separately with a gray line, with the median log number of sequences per week shown in a thicker red line. The log of the total data per week is shown in blue.}
\label{plot::data_over_time}
\end{center}
\end{figure}

Several data quality issues complicate the use of models for nowcasting and forecasting variant proportions across locations.
One data quality issue is delayed reporting. Variant identification requires a viral specimen, such as a nasal swab, which then must be analyzed in a laboratory. Because the specimen collection, testing, and sequencing process takes time, obtaining complete real-time data is impossible. This leads to the need for models that can nowcast or forecast with reasonable accuracy using incomplete data, with recent data being most limited \citep{Nixon2022}.
Another data quality issue is the quantity of sequences analyzed, which, for COVID-19, has decreased dramatically over time (Figure \ref{plot::data_over_time}). This means that models have to create predictions with very sparse data. In 2022, there was an average of approximately 2,900 daily sequences reported across the United States. In 2024 up to November 4th, which is the last date we have data in that year when the data for this paper were finalized, there were approximately 265 daily sequences collected. Over just two years, sequences have dropped by an order of magnitude.       
Data quantity issues are magnified when attempting to make predictions at resolutions below the country level, e.g., at the state level in the US. That is because there is less data at these resolutions, and the data quantity and availability can vary dramatically by state (Figure \ref{plot::data_over_time}). Lower counts of data can lead to noisier proportions which can be harder to model.

Multinomial logistic regression (MLR) is often used to model SARS-CoV-2 variants \citep{annavajhala2021emergence, faria2021genomics}. These models assume the counts of the variants follow a multinomial distribution with probabilities that are often modeled as linear in time on the logit scale. To account for the differences across locations, hierarchical multinomial logistic regression (HMLR) can be used to allow for data sharing, which is an approach that has been used in the literature to model variants at the country level \citep{Susswein2023,Abousamra2024}.
In this paper, we create a taxonomy for HMLR models that have been used to predict SARS-CoV-2 variant prevalences, review existing HMLR models in the literature, introduce several HMLR model variations, and compare their performance on retrospective datasets. 

We split this paper into the following sections.
Section 2 gives a brief introduction to genomic SARS-CoV-2 data and how we created the retrospective datasets.
Section 3 gives an overview of MLR and HMLR models and how they have previously been used in the literature, and introduces a framework for a general class of HMLR models.
Section 4 contains a motivating example applying the framework to modeling state-level SARS-CoV-2 variant prevalences and evaluating models' performance. 
Section 5 provides a brief discussion on the HMLR models and possible next steps.

\section{Data}

Comparing nowcast and forecast performance of multiple models for SARS-CoV-2 variant prevalences requires careful construction of training and evaluation data that reflect the availability of SARS-CoV-2 genomic sequence data in real time. One specific feature of the data pipeline introduces challenges for model evaluation: delays of two weeks or more between specimen collection and reporting are common, making it important to have time-stamped ``versions'' of which sequences were available for analysis on a given date. In what follows, we describe how SARS-CoV-2 genomic sequence data are obtained with a focus on why delays occur. We also describe how we formulate the nowcasting and forecasting problem and construct the corresponding datasets.

\subsection{SARS-CoV-2 Sequencing}
The process for obtaining SARS-CoV-2 genomic sequence data that provides detailed information about which pathogen variant infected an individual is time-consuming and expensive. 
Genomic sequence data are generated by first taking a specimen from an infected host, typically a nasal swab. Next, the specimen is inactivated, and the nucleic acid is extracted. This extract is then tested for the presence of SARS-CoV-2 using a Polymerase Chain Reaction (PCR) test. If the test is positive, the extracted RNA is then sequenced \citep{WHO2021Genomic}.   
These data result in a SARS-CoV-2 genomic sequence, approximately 29 thousand base-pairs, assigned to that individual's infection \citep{raskin2021genetics}.
The sequencing process is often conducted in batches, where each batch corresponds to a single PCR plate with a fixed number of wells, each of which can be filled with a single specimen. Often, specimens will be stored until all wells on a single plate can be filled. Depending on the volume of specimens collected, this process can cause delays of weeks or months between specimen collection and the availability of variant counts. 

An additional layer of data processing makes these data available for population-level modeling.
Most SARS-CoV-2 sequences from specimens collected in the United States are stored in the GenBank database, which is the publicly available genetic sequence database of the National Institutes of Health (NIH) \citep{benson2012genbank}.
External research groups, such as the Nextstrain team and others, run algorithms on the full datasets that estimate a likely evolutionary tree based on the sequence data, which leads to a variant nomenclature model \citep{hadfield2018nextstrain}. 
These evolutionary trees are re-estimated periodically, usually every month or two.
One byproduct of the evolutionary tree model is that each sequence is labeled as belonging to a class of related sequences. 
There have been three main nomenclatures proposed for SARS-CoV-2 variants that attempt to aggregate and classify these base pair mutations: Pango lineages, \citep{rambaut2020dynamic}, GISAID clades, \citep{GISAID2021}, and Nextstrain clades \citep{aksamentov2021nextclade}. 
Nextstrain clades and GISAID clades describe large-scale trends in SARS-CoV-2 evolution and provide broader classifications than finer-grained classifications from Pango lineages. 

To facilitate reproducible modeling with a smaller number of genomic groups, we decided to use the Nextstrain clades as the classification approach for our modeling, consistent with the choice made by the SARS-CoV-2 Variant Nowcast Hub \citep{MacArthur2026CollaborativeNowcasting}. The terms ``variant'' and ``clade'' do not have one agreed-upon definition; for the purposes of this paper, we will let ``variant'' refer to a general lineage of an infectious disease, whereas ``clade'' will refer to a Nextstrain clade, a more specific unit. We will use variant to refer to what we are attempting to forecast and nowcast, and only use clade when referring to a particular Nextstrain-defined classification.    
The data used in this project are public, courtesy of GenBank and Nextstrain; Nextstrain curates a public ``full open metadata'' dataset where all current genomic sequence data from Genbank are made available in a tabular format \citep{hadfield2018nextstrain}. 
These files contain information on the viral sequences, including what Nextstrain clade was assigned to the sequence, what date the sequence was collected, where the case occurred, and when the sequence was submitted to the database.

\subsection{Problem Formulation and Dataset Construction}\label{sec::dataset}

The United States SARS-CoV-2 Variant Nowcast Hub motivates our definition of the nowcasting and forecasting problem and the construction of corresponding datasets.
This modeling hub curates nowcasts and forecasts of the proportion of at most ten of the most common SARS-CoV-2 clades for the fifty US states as well as Puerto Rico and Washington D.C. \citep{MacArthur2026CollaborativeNowcasting}. Since October 2024, these nowcasts and forecasts have been made every Wednesday at the daily resolution with a nowcast period of 32 days, which includes the submission date and the 31 days before that date, and a forecast horizon of 10 days after the submission date. 

The experimental validation of models described in this work used aggregate daily counts of sequenced specimens from the time period of January 1, 2022 through November 22, 2024. For each sequenced specimen, our dataset includes the date the specimen was collected and the date when sequencing data was reported.
We exclude sequences from non-human hosts and those from locations other than the US states, Puerto Rico, and Washington D.C. 
We fixed the the probabilistic clade assignment model that was in operation on November 22, 2024 and used this to assign clades retrospectively to all sequence data.
Fixing the probabilistic clade assignment model leads to a difference from what would be seen if this analysis were done in real time. The dataset was constructed this way because while in real-time it is possible to obtain the data necessary to reconstruct past trees and data, this data was not available for the years in the past needed for this analysis. See section \ref{sec::Disc} for more discussion. 

\begin{figure}[ht]
\begin{center}
\includegraphics[width=0.9\columnwidth,height=0.8\textheight, keepaspectratio]{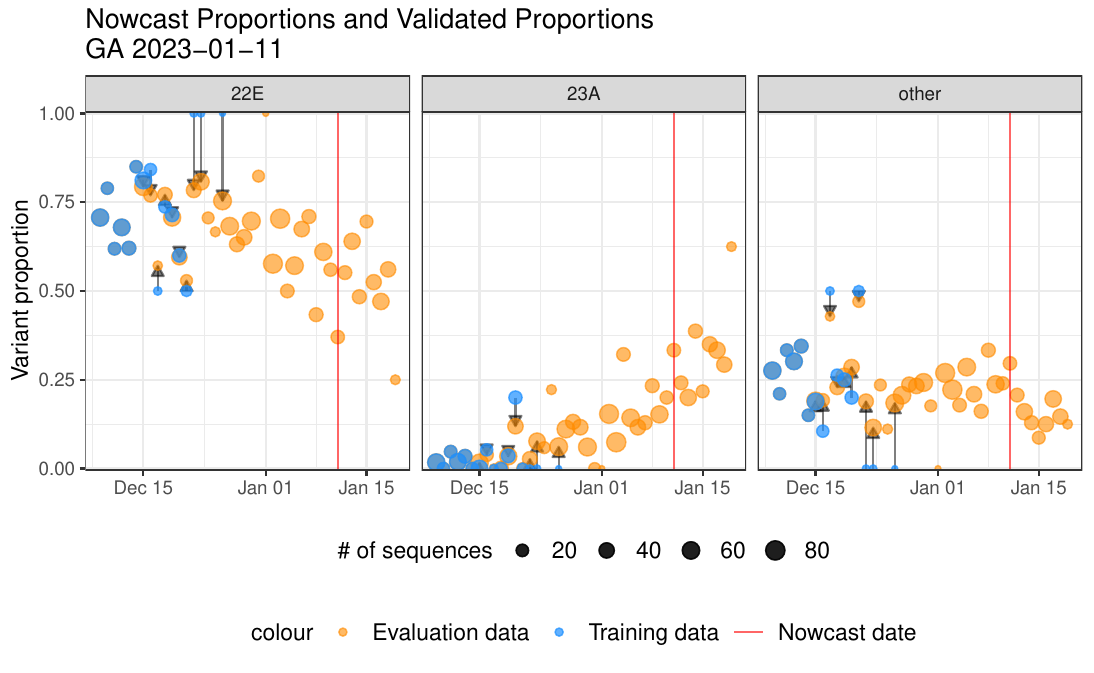}
\caption{A plot showing the proportions of the clades 22E and 23A, and, for illustrative purposes, all other clades grouped into others, for Georgia (labeled GA). The blue dots are the clade proportions available for modeling on a submission date of 2023-01-11 for the days we are interested in nowcasting, and the orange dots are the data available 91 days later. If there was data available for a date on 2023-01-11 and 91 days later, and the proportion changed, that change is denoted by an arrow. The red line denotes the submission date 2023-01-11. }
\label{plot::Arrow_plot_GA}
\end{center}
\end{figure}

We define 106 weekly submission dates from Wednesday, August 3, 2022 to Wednesday, August 7, 2024. Previous analyses suggest that most sequencing data is reported within 91 days of the specimen collection date  \citep{MacArthur2026CollaborativeNowcasting}. Keeping all the sequencing data collected up to 150 days before the submission date allows for a reasonable amount of training data, even with long delays in sequencing. Thus, we created 119 datasets for dates from Wednesday, August 3, 2022 to November 6, 2024. Each dataset contained all data reported by the submission date, with a collection date of at most 150 days before the submission date. To train a model to make predictions for a submission date, we fit the model to all data from the dataset associated with that submission date. Each submission is evaluated using the data from the dataset with a date 91 days after the submission date, e.g. the dataset for November 6, 2024 is used to evaluate the submissions for August 7, 2024. Datasets constructed to train and evaluate models for a single submission date and two selected clades are illustrated in Figure \ref{plot::Arrow_plot_GA}. We can see that data availability lags can obscure emerging trends. This is common in the data and important to preserve in the retrospective datasets.

\begin{figure}[ht]
\begin{center}
\includegraphics[width=0.9\columnwidth,height=0.8\textheight, keepaspectratio]{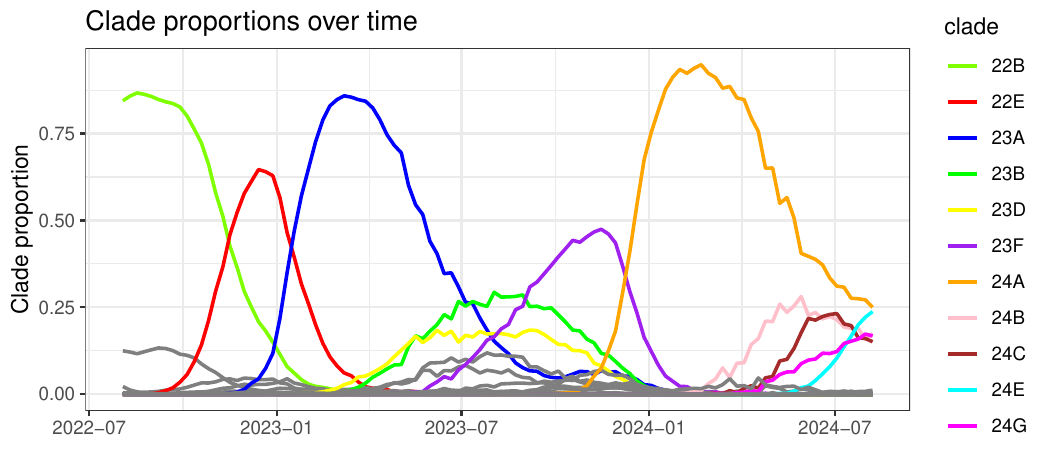}
\caption{A plot showing the proportions of the clades at the national level; all clades that ever reached a national proportion of 0.15 are colored.}
\label{plot::clades_over_time}
\end{center}
\end{figure}

To choose what clades were modeled on a given week, we identify up to nine clades that account for at least one percent of the national average of cases in at least one of the previous three complete MMWR weeks \citep{MMWR} prior to the Wednesday of the week of the submission date, in line with the criteria used by the Variant Nowcast Hub \citep{MacArthur2026CollaborativeNowcasting}. All remaining clades are combined into a category labeled ``other''. If there are more than nine clades that satisfy these criteria, the nine most prevalent are chosen. Because of the nature of SARS-CoV-2 evolution, there are both periods with one dominant clade and many competing clades (Figure \ref{plot::clades_over_time}). 

\section{Models for Multinomial Data}

\subsection{Multinomial Logistic Regression }\label{sec::mlr}

% introduction to notation
Let $n^{(s)}_{lt}$ refer to the total number of viral specimens collected from a location $l$ at time $t$ reported as of time $s$, $c^{(s)}_{ltv}$ refer to the number of viral specimens associated with variant $v$ from location $l$ at time $t$, $\boldsymbol c^{(s)}_{lt} = (c^{(s)}_{lt1}, \dots, c^{(s)}_{ltV})$ refer to the vector of counts of genomic sequences associated with $V$ variants, and $\boldsymbol x_{lt} = (x_{lt1}, \dots, x_{ltP})$  be a $P \times 1$ vector of covariates at location $l$ and time $t$.

% introduction to MLR
To model the number of viral specimens per variant at time $t$ in location $l$ reported as of time $s$, we assume that these counts follow a multinomial distribution with total $n^{(s)}_{lt} = \sum_{v = 1}^V c^{(s)}_{ltv}$ and unknown probabilities 
$\theta_{ltv}/\sum_{v = 1}^V \theta_{ltv}$ for $v = 1, \dots, V$. Letting  $\boldsymbol \theta_{lt} = (\theta_{lt1}, \ldots, \theta_{ltV})$, we assume
\begin{align}
\boldsymbol c^{(s)}_{lt}  \mid n^{(s)}_{lt}, \boldsymbol \theta_{lt}&\stackrel{indep.}{\sim} \text{Multinomial}\left(n^{(s)}_{lt}, \left(\frac{\theta_{lt1}}{\sum_{v = 1}^V \theta_{ltv}}, \dots,\frac{\theta_{ltV}}{\sum_{v = 1}^V \theta_{ltv}}\right)\right).\label{eq:mnreg}
\end{align}
The multinomial probabilities in turn depend on a set of observed covariates $\boldsymbol x_{lt}$ and a $P\times 1$ vector of unknown regression coefficients $(\gamma_{lv1}, \dots, \gamma_{lvP})$ for location $l$ and variant $v$.
One approach to fit this model assumes a deterministic log-linear relationship between $\theta_{ltv}$ and the covariates:
\begin{align}
\theta_{ltv}  &= \text{exp}\left\{\sum_{p=1}^P x_{ltp} \gamma_{lvp} \right\} \text{ for } v=1,\dots, V.\nonumber
\end{align}
When the above relationship is assumed, then the model in (\ref{eq:mnreg}) is a multinomial logistic regression. An alternate approach is to treat the $\theta_{ltv}$ as independent random variables 
\begin{align}
\theta_{ltv}|\gamma_{lv1}, \dots, \gamma_{lvP}\sim\text{Gamma}(\text{shape}=\text{exp}\{\sum_{p=1}^P x_{ltp} \gamma_{lvp} \}, \text{rate} = 1)\nonumber
\end{align}
for $v = 1, \dots, V$.
With this additional probabilistic structure, the model becomes a Dirichlet-Multinomial logistic regression model which assumes that counts $\boldsymbol c^{(s)}_{lt}$ are conditionally Dirichlet-Multinomial distributed given the coefficients $\gamma_{lv1}, \dots, \gamma_{lvP}$ with probabilities proportional to $\mathbb{E}\left[\theta_{ltv} | \gamma_{lv1}, \dots, \gamma_{lvP}\right] = \text{exp}\left\{\sum_{p=1}^P x_{ltp} \gamma_{lvp} \right\}$.
This allows for overdispersion of the variant counts $c^{(s)}_{ltv}$ and approximates the multinomial logistic regression model as $\mathbb{E}[\theta_{ltv}|\gamma_{lv1}, \dots, \gamma_{lvP}] \rightarrow \infty$ while holding the probabilities $\left(\frac{\theta_{lt1}}{\sum_{v = 1}^V \theta_{ltv}}, \dots,\frac{\theta_{ltV}}{\sum_{v = 1}^V \theta_{ltv}}\right)$ constant for all $v$.

% parameter constraints to make model identifiable
Given the choice of a multinomial or Dirichlet-Multinomial logistic regression model for variant counts $c^{(s)}_{ltv}$, constraints must be imposed on the coefficients $(\gamma_{lv1}, \dots, \gamma_{lvP})$ to ensure that they are identifiable. Because of the constraint that the $\theta_{ltv}$ must sum to one, this model is over-parameterized in that $\theta_{ltV}$ can be reconstructed from  $\theta_{lt1},\dots, \theta_{lt(V-1)}$. 
For identification of $(\gamma_{lv1}, \dots, \gamma_{lvP})$, a variant $v$ is chosen as a reference with a $\gamma_{lv1} = \dots = \gamma_{lvP} = 0$. For this paper, we have chosen to set the last variant as a reference, $\gamma_{lV1} = \dots = \gamma_{lVP} = 0$. Because of the way that this data was structured as described in Section~\ref{sec::dataset}, this results in the clade labeled ``other'' being the reference variant. A consequence of choosing this reference variant to have $\gamma_{lV1} = \dots = \gamma_{lVP} = 0$ is that this allows coefficients $\gamma_{lvr}$ for $v < V$ to be interpreted as the relative change in log odds of variant $v$  to reference variant $V$ associated with a one unit change in $x_{ltr}$, holding all else constant. This is because this model is equivalent to assuming that $\log(\theta_{ltv}/\theta_{ltV}) = \sum_{p=1}^P x_{ltp} \gamma_{lvp} \text{ for } v=1,\dots, V-1.\nonumber$. 

For the modeling of SARS-CoV-2 variants using multinomial logistic regression, it is often assumed that the covariates are made up of an intercept and a time trend, i.e. $P = 2$, with $x_{lt1} = 1$, and $x_{lt2} = t$  \citep{Susswein2023,Abousamra2024, annavajhala2021emergence}. This model allows the time coefficients $\gamma_{lv2}$ to be interpreted as relative growth advantages over time compared to the reference variant \citep{Susswein2023}. That is if $\gamma_{lv2}< 0$, we can say that variant $v$ will grow more slowly than the reference variant over one unit of time in location $l$, and if $\gamma_{lv2} > 0$, we can say that variant $v$ is growing more rapidly than the reference variant over one unit of time in location $l$.

In what follows, we recognize that the assumption of a log-linear time trend can be unrealistic and restrictive and consider the inclusion of additional covariates beyond an intercept and a time trend. For ease of comparison to other implementations of multinomial logistic regression models for nowcasting and forecasting variant prevalences and to restrict the scope of considered models to models with polynomial time trends we assume that $P \geq 2$ with $x_{ltp} = t^{p-1}$. Practically, this means that all models we consider have at least an intercept and a linear time trend. Under this constraint on the covariates $x_{ltp}$, the covariate values do not depend on location. In what follows, we drop the $l$ subscript and refer to $x_{ltp}$ as $x_{tp}$ and $\boldsymbol x_{lt}$ as $\boldsymbol x_t$.

\subsection{Hierarchical Multinomial Logistic Regression}
\subsubsection{HMLR Framework}\label{sec::hmlr}
When using multinomial or Dirichlet-Multinomial logistic regression models to nowcast and forecast variant prevalences across multiple locations and/or with limited data, a hierarchical model that shares information across locations can improve model accuracy and stability \citep{Susswein2023, Abousamra2024}. 
The hierarchical structure that we consider assumes that vectors of regression coefficients corresponding to the same predictor (for example, the slope of the log-linear relationship with time) across locations share a common mean and covariance structure. This allows for information sharing across locations, which can improve estimation of coefficients in locations with little or no data. Equivalently, the hierarhical structure makes the mecahnistic assumption that a variant's relative growth advantage in one location is likely to be similar in another location. This is appropriate and beneficial when the same coefficients across locations are, in fact, similar.
For the purposes of this paper, we assume the independence of regression coefficients across locations and predictors, e.g, we assume that the intercept and time trend coefficients are independent. Allowing for correlated regression coefficients across locations and predictors is a possible direction for future work discussed in Section \ref{sec::Disc}. 

To describe our assumed HMLR structure, we let $\boldsymbol \gamma_{lv\cdot}$ refer to the $P\times 1$ vector of coefficients for all predictors at location $l$ for variant $v$, e.g all of the time polynomial terms for one variant at one location. Let $\boldsymbol \gamma_{l\cdot p}$ refer to the $\left(V-1\right)\times 1$ vector of coefficients for each non-reference variant at location $l$ and for predictor $p$ e.g a linear time coefficient across variants, $v_1,\ldots, V-1$.
Then our HMLR models assume prior distributions for $\boldsymbol \gamma_{l \cdot p}$, the $\left(V-1\right)\times 1$ vectors of regression coefficients for location $l$ and predictor $p$, 
\begin{align}
    \boldsymbol \gamma_{l\cdot p} | \boldsymbol \mu_{p}, \boldsymbol \Sigma_{p} \stackrel{i.i.d.}{\sim} \text{normal}\left(\boldsymbol \mu_{p}, \boldsymbol \Sigma_{p} \right) \text{ for } l = 1, \dots, L \text{ and } p = 1, \dots, P, \label{eq:hier}
\end{align}  
where regression coefficients $\boldsymbol \gamma_{l \cdot p}$ are assumed to be independent across locations $l$ and regression coefficients $p$. Here $\boldsymbol \mu_p$ refers to a $\left(V-1\right)\times 1$ overall mean vector for each regression coefficient across non-reference variants, and $\boldsymbol \Sigma_p$ refers to a $\left(V-1\right)\times \left(V-1\right)$ overall covariance matrix for each regression coefficient across non-reference variants. We discuss types of structures for $\boldsymbol \Sigma_p$ in the following sections.

In summary, our HMLR model for variant counts $\boldsymbol c^{(s)}_{lt}$ at locations $l = 1, \dots, L$ and times $t = 1, \dots, T$ reported as of time $s$ as defined by ~\eqref{eq:mnreg} and ~\eqref{eq:hier} assumes:
\begin{eqnarray}
\boldsymbol c^{(s)}_{lt}  \mid n^{(s)}_{lt}, \boldsymbol \theta_{lt} & \stackrel{indep.}{\sim} & \text{Multinomial}\left(n^{(s)}_{lt}, \left(\frac{\theta_{lt1}}{1 + \sum_{v = 1}^{V-1} \theta_{ltv}}, \dots,\frac{\theta_{lt(V-1)}}{1 + \sum_{v = 1}^{V-1} \theta_{ltv}}, \frac{1}{1 + \sum_{v = 1}^{V-1} \theta_{ltv}}\right)\right) \label{eq:framework-count} \\ 
\mathbb{E}\left[\theta_{ltv} | \boldsymbol \gamma_{lv\cdot}\right] & = & \text{exp}\left\{\boldsymbol \gamma_{lv\cdot}'\boldsymbol x_{t} \right\} \text{ for } v=1,\dots, V-1,\label{eq:framework-mean} \\
\boldsymbol \gamma_{l\cdot p} | \boldsymbol \mu_{p}, \boldsymbol \Sigma_{p} & \stackrel{i.i.d.}{\sim} & \text{normal}\left(\boldsymbol \mu_{p}, \boldsymbol \Sigma_{p} \right) \text{ for } l = 1, \dots, L \text{ and } p = 1, \dots, P,\label{eq:framework-hierarchy}
\end{eqnarray}
To be used in practice, an HMLR model requires further specification of values or prior distributions for $\boldsymbol \mu_p$ and $\boldsymbol \Sigma_p$ for $p = 1, \dots, P$. In the next sections, we review special cases of this HMLR model, including specification of values or prior distributions for $\boldsymbol \mu_p$ and $\boldsymbol \Sigma_p$, that have been used in related literature, and then characterize a more general taxonomy of HMLR models that apply to the state-level variant proportion nowcasting and forecasting problem.
\subsubsection{HMLR models used in practice}

The general HMLR model described in equations \ref{eq:framework-count}, \ref{eq:framework-mean}, and \ref{eq:framework-hierarchy} generalizes the models used in Susswein et al. \citep{Susswein2023} and Abousamra et al. \citep{Abousamra2024}. Both Susswein et al. \citep{Susswein2023} and Abousamra et al. \citep{Abousamra2024} find HMLR models to be well-suited to modeling the SARS-CoV-2 variant dynamics and for estimating relative growth advantages across variants. Specifically, Abousamra et al. \citep{Abousamra2024} finds that HMLR models outperform their non-hierarchical MLR counterparts when being used to predict the short-term behavior (30 days in the future) of SARS-CoV-2 variants at the country level in countries with less available data.  
We summarize the modeling choices made in Susswein et al. \citep{Susswein2023} and Abousamra et al. \citep{Abousamra2024} in Table~\ref{tab:otherheir}, using $\boldsymbol 0_V$ to refer to an $V\times 1$ vector with all elements equal to $0$, $\boldsymbol I_V$ to refer to a diagonal $V\times V$ matrix with all diagonal elements equal to $1$, and $\text{diag}\left\{\boldsymbol \sigma^2_v \right\}$ to refer to a diagonal matrix with diagonal elements equal to $\boldsymbol \sigma^2_v $ and all other elements equal to $0$ .

\begin{table}[h]
\centering
\begin{tabular}{ll}
 Abousamra et al. \citep{Abousamra2024}  & Susswein et al. \citep{Susswein2023} \\ \hline \hline
 $\theta_{ltv}=\text{exp}\{ \gamma_{lv1} +  \gamma_{lv2}t\}$ & $\theta_{ltv}=\text{exp}\{ \gamma_{lv1} +  \gamma_{lv2}t\}$\\ \hline
  $\boldsymbol \mu_{1} =\boldsymbol 0_{V-1}$ &  $\mu_{1j} \stackrel{i.i.d.}{\sim} t(-5,5,3) $ \\ \hline
 $\boldsymbol \Sigma_{1} = 6 \boldsymbol I_{V-1}$ & $\boldsymbol \Sigma_{1} = \text{diag}\left\{\boldsymbol \sigma^2_v \right\}$, ${\sigma^2_{vj}}\stackrel{i.i.d.}{\sim}  \text{normal}_+\left(2, 1\right)$ \\ \hline
\multirow{2}{*}{$\boldsymbol \mu_{2}\sim \text{normal}\left(\boldsymbol 0, 0.2^2\boldsymbol I_{V-1} \right)$} & $\boldsymbol \mu_{2}|\mu_h, \sigma^2_h \sim \text{normal}\left(\mu_{h}\boldsymbol 1_{V-1}, \sigma^2_h \boldsymbol I_{V-1}\right)$ \\
& $\mu_h\sim \text{normal}(-1, 0.5)$, $\sigma^2_h\sim \text{normal}(0.1, 1)$ \\ \hline
\multirow{2}{*}{$\boldsymbol \Sigma_{2} = \sigma^2 \boldsymbol I_{V-1}$, $\sigma^2\sim \text{normal}_+\left(1, 0.1\right)$} &  $\boldsymbol \Sigma_{2} = \text{diag}\left\{\boldsymbol \sigma^2 \right\}\boldsymbol \Omega \text{diag}\left\{\boldsymbol \sigma^2 \right\}$,  \\
 & $\sigma^2_j \stackrel{i.i.d.}{\sim} \text{normal}(0.5, 2), \boldsymbol \Omega \sim \text{LKJ}(2)  $\\ \hline 
\end{tabular}
\caption{The hierarchical structures of the models from Abousamra et al.\citep{Abousamra2024} and  Susswein et al.\citep{Susswein2023}}
\label{tab:otherheir}
\end{table}

The HMLR models implemented in Susswein et al. \citep{Susswein2023} and Abousamra et al. \citep{Abousamra2024} share the same treatment of the unnormalized variant prevalences $\theta_{ltv}$ and how they vary over time. Both treat $\theta_{ltv}$ as fixed and equal to $\text{exp}\{\boldsymbol \gamma_{lv\cdot}'\boldsymbol x_t \}$, which corresponds to assuming multinomial distributed variant counts. Both also assume that the unnormalized variant prevalences $\theta_{lvt}$ are log-linear in time with $P = 2$. They also both treat the variant with the greatest amount of data as the reference variant, $\theta_{ltV}$.

The models differ in their treatment of the prior means $\boldsymbol \mu_p$ and variances $\boldsymbol \Sigma_p$ for the $p = 1, \dots, P$ regression coefficients. Abousamra et al. \citep{Abousamra2024} fixes $\boldsymbol \mu_1$, whereas Susswein et al. \citep{Susswein2023} assumes prior distributions for elements of $\boldsymbol \mu_1$ with fixed prior parameters. Abousamra et al. \citep{Abousamra2024} assume a prior distribution with fixed prior parameters for elements of $\boldsymbol \mu_2$, whereas Susswein et al. \citep{Susswein2023} assume a prior distribution with varying prior parameters $\mu_h$ and $\sigma^2_h$ for elements of $\boldsymbol \mu_2$, and furthermore assume prior distributions with fixed parameters for $\mu_h$ and $\sigma^2_h$. We can think of these different choices of priors for $\boldsymbol \mu_p$ as representing the choice of the number of levels in a hierarchical model; fixing $\boldsymbol \mu_p$ corresponds to a one level hierarchical model, assuming a prior with fixed parameters for $\boldsymbol \mu_p$ corresponds to a two level hierarchical model, and assuming priors with varying parameters that are shared across variants for $\boldsymbol \mu_p$ corresponds to a three level hierarchical model. Including at least two levels allows for information sharing across variants through estimation of the regression coefficient means $\boldsymbol \mu_p$ and covariance matrices $\boldsymbol \Sigma_p$ for $p = 1, \dots, P$.

In addition to considering different levels of hierarchy for regression coefficient prior covariance matrices $\boldsymbol \Sigma_p$ for $p = 1, \dots, P$, Abousamra et al. \citep{Abousamra2024} and Susswein et al. \citep{Susswein2023} differ in their assumptions on the structure of the covariance matrices. Abousamra et al. \citep{Abousamra2024} assume that the prior covariance for intercepts and slopes across variants, $\boldsymbol \Sigma_1$ and $\boldsymbol \Sigma_2$, are proportional to identity matrices. This reflects an assumption that intercepts and slopes are independent and equally variable across variants. In contrast, Susswein et al. \citep{Susswein2023} assume that the intercept covariance matrix $\boldsymbol \Sigma_1$ is diagonal and the slope covariance $\boldsymbol \Sigma_2$ matrix is unstructured. This reflects an assumption that intercepts are independent across variants but that some variants have more or less variable intercepts across locations, and an assumption that some variants have not only more or less variable slopes across locations but also that some pairs of variants tend to have more similar slopes than other pairs.

\subsection{Modeling SARS-CoV-2 clade data using HMLR}\label{sec::hmlrused}

Our review of HMLR models used in Susswein et al. \citep{Susswein2023}  and  Abousamra et al. \citep{Abousamra2024} highlights some ways in which HMLR models may vary: the number of levels of hierarchy used when modeling the regression coefficient means $\boldsymbol \mu_p$ and covariance matrices $\boldsymbol \Sigma_p$ and the structures assumed for the regression coefficient covariances matrices $\boldsymbol \Sigma_p$. In the context of the more general framework for HMLR models we defined in Section~\ref{sec::hmlr}, additional choices that need to be made when implementing HMLR models are apparent. Although Susswein et al. \citep{Susswein2023} and Abousamra et al. \citep{Abousamra2024} both assume linear time trends with $P = 2$, $x_{t1} = 1$, and $x_{t2} = t$, higher order polynomial time trends with $P \geq 2$ can be used, especially when estimation of relative growth advantages is not of primary interest. Additionally, although Susswein et al. \citep{Susswein2023} and Abousamra et al. \citep{Abousamra2024} both assume multinomial variant counts, Dirichlet-Multinomial variant counts can alternatively be assumed, especially when variant counts are overdispersed.

\begin{table}[h]
\centering
\begin{tabular}{l|l|l|l} 
\multicolumn{4}{l}{Model Components} \\ \hline \hline
\multirow{3}{*}{Common across models} &\multicolumn{3}{l}{$\boldsymbol c^{(s)}_{lt}  \mid n^{(s)}_{lt}, \boldsymbol \theta_{lt} \stackrel{indep.}{\sim} \text{Multinomial}\left(n^{(s)}_{lt}, \left(\frac{\theta_{lt1}}{1 + \sum_{v = 1}^{V-1} \theta_{ltv}}, \dots,\frac{\theta_{lt(V-1)}}{1 + \sum_{v = 1}^{V-1} \theta_{ltv}}, \frac{1}{1 + \sum_{v = 1}^{V-1} \theta_{ltv}}\right)\right)$} \\ 
 &\multicolumn{3}{l}{$\mathbb{E}\left[\theta_{ltv} | \boldsymbol \gamma_{lv\cdot}\right] = \text{exp}\left\{\boldsymbol \gamma_{lv\cdot}'\boldsymbol x_{t} \right\}$, \quad $x_{tp} = t^{p-1}$,\quad $\boldsymbol \gamma_{l\cdot p} | \boldsymbol \mu_{p}, \boldsymbol \Sigma_{p}  \stackrel{i.i.d.}{\sim} \text{normal}\left(\boldsymbol \mu_{p}, \boldsymbol \Sigma_{p} \right)$} \\
 &\multicolumn{3}{l}{$\mu_{pj} \stackrel{i.i.d.}{\sim} \text{Normal}(0, 400^2)$} \\ \hline
\multirow{11}{*}{Varying across models} & \multirow{7}{*}{Covariance} & \multirow{2}{*}{\underline{S}hared} & $\boldsymbol\Sigma_p = \sigma_p^2 \boldsymbol I_{V-1}$ \\ 
& & & $\sigma_p^2 \sim \text{N}_+(1, 400^2)$ \\ \cline{3-4}
& & 
 \multirow{2}{*}{\underline{I}ndividual} & $\boldsymbol\Sigma_p = \text{diag}\{\boldsymbol\sigma_p^2\}$ \\
  & &  & $\sigma_{pj}^2 \stackrel{i.i.d.}{\sim} \text{N}_+(1, 400^2)$\\ \cline{3-4}
& & \multirow{3}{*}{\underline{C}orrelated} & $\boldsymbol\Sigma_p = \text{diag}\{\boldsymbol \sigma_p \} \boldsymbol \Omega_p \text{diag}\{\boldsymbol \sigma_p \}$ \\ 
& & & $\sigma_{pj}^2 \stackrel{i.i.d.}{\sim} \text{N}_+(1, 400^2)$ \\ 
& & & $\boldsymbol \Omega_p \sim \text{LKJ}(2)$ \\ 
\cline{2-4} 
& \multirow{2}{*}{Polynomial} & \underline{L}inear & $P = 2$\\
& & \underline{C}ubic & $P = 4$ \\ \cline{2-4}
 & \multirow{2}{*}{Data} & \underline{M}ultinomial & $\theta_{ltv} = \exp\!\left\{ \boldsymbol\gamma_{lv\cdot}' \boldsymbol x_t \right\}$\\
 & & \underline{D}irichlet-Multinomial & $\theta_{ltv} \sim \text{Gamma}\!\left(
 \exp\!\left\{ \boldsymbol\gamma_{lv\cdot}' \boldsymbol x_t \right\},
1
\right)$ \\ \hline \hline
\multicolumn{4}{c}{} \\
\multicolumn{4}{l}{Abbreviations} \\ \hline \hline
Covariance & Polynomial & Data & Abbreviation \\ \hline
\multirow{4}{*}{\underline{S}hared}  & \multirow{2}{*}{\underline{L}inear} & \underline{M}ultinomial & SLM \\ \cline{3-4}
 &   & \underline{D}irichlet-Multinomial & SLD \\ \cline{2-4}
  & \multirow{2}{*}{\underline{C}ubic} & \underline{M}ultinomial & SCM \\ \cline{3-4}
 &   & \underline{D}irichlet-Multinomial & SCD \\ \cline{1-4}
 \multirow{4}{*}{\underline{I}ndividual}  & \multirow{2}{*}{\underline{L}inear} & \underline{M}ultinomial & ILM \\ \cline{3-4}
 &   & \underline{D}irichlet-Multinomial & ILD \\ \cline{2-4}
  & \multirow{2}{*}{\underline{C}ubic} & \underline{M}ultinomial & ICM \\ \cline{3-4}
 &   & \underline{D}irichlet-Multinomial & ICD \\ \cline{1-4}
 \multirow{4}{*}{\underline{C}orrelated}  & \multirow{2}{*}{\underline{L}inear} & \underline{M}ultinomial & CLM \\ \cline{3-4}
 &   & \underline{D}irichlet-Multinomial & CLD \\ \cline{2-4}
  & \multirow{2}{*}{\underline{C}ubic} & \underline{M}ultinomial & CCM \\ \cline{3-4}
 &   & \underline{D}irichlet-Multinomial & CCD \\ \hline
 \end{tabular}
 \caption{Hierarchical multinomial logistic regression (HMLR) models considered and abbreviations. The model most similar to Abousamra et al. \citep{Abousamra2024} is SLM, and the model most similar to Susswein et al. \citep{Susswein2023} is CLM.}
 \label{tab::HMLR_models}
\end{table}

In order to systematically examine how different choices made in the formulation of HMLR models affect their nowcasting and forecasting performance in the context of variant prevalences, we define a set of 12 HMLR models that assume varying regression coefficient covariance structures, time trends, and variant count distributions. These models are summarized in Table~\ref{tab::HMLR_models}. Because we apply these HMLR models to nowcast and forecast variant prevalences at the state level and some states have limited or no data at certain times, all HMLR models we consider assume two-level hierarchical models for the regression coefficient means $\boldsymbol \mu_p$ and covariance matrices $\boldsymbol \Sigma_p$ that facilitate information sharing across states. For all 12 models, we assume a diffuse normal prior with mean $0$ and variance $400^2$ for elements of the regression coefficient means $\boldsymbol \mu_p$. 

We vary the assumed distribution of the variant counts to be multinomial (M) with fixed $\theta_{ltv}$ given the regression coefficients $\boldsymbol \gamma_{lv\cdot}$ or Dirichlet-Multinomial (D) with gamma distributed $\theta_{ltv}$ given the regression coefficients $\boldsymbol \gamma_{lv\cdot}$, the degree of the polynomial time trend to be linear (L) with $P = 2$ or cubic (C) with $P = 4$, and the structure of the regression coefficient covariances $\boldsymbol \Sigma_p$ to be shared (S) with $\boldsymbol \Sigma_p = \sigma^2 \boldsymbol I_{V-1}$, individually varying across variants (I) with $\boldsymbol \Sigma_p = \text{diag}\{\boldsymbol \sigma^2_p\}$, or correlated across variants (C) with $\boldsymbol \Sigma_p =  \text{diag}\{\boldsymbol \sigma_p \} \boldsymbol \Omega_p \text{diag}\{\boldsymbol \sigma_p \}$, where $\boldsymbol \Omega_p$ is a $\left(V-1\right)\times \left(V-1\right)$ correlation matrix and $\sigma_{p}$ is a $V-1$ length vector of standard deviations. For all variances, we assume a diffuse half-normal prior with mean $1$ and variance $400^2$. As in Susswein et al. \citep{Susswein2023}, we assume an LKJ$\left(2\right)$ prior for each correlation matrix $\boldsymbol \Omega_p$   \citep{Lewandowski2009}. Models are given three-letter abbreviations that describe the covariance structure assumed for $\boldsymbol \Sigma_p$, the assumed polynomial degree of the time trend, and the assumed distribution of the data conditional on the regression coefficients. The model most similar to Abousamra et al. \citep{Abousamra2024} is SLM, and the model most similar to Susswein et al. \citep{Susswein2023} is CLM.

\subsection{Forecasting Setup}

We fit all 12 HMLR models described in Section~\ref{sec::hmlrused} and a baseline MLR model to the 106 datasets corresponding to the 106 submission dates from Wednesday, August 3, 2022 to Wednesday, August 7, 2024 described in Section~\ref{sec::dataset}. Each dataset used to fit a model corresponds to a distinct submission date and contains all sequencing data collected within 150 days of the submission date and reported as of the submission date. This yields in 150 days of training data per submission date.  For each submission date and corresponding dataset, clades for nowcasting and forecasting are determined as described in Section~\ref{sec::dataset}.

A baseline MLR model allows us to assess the value of HMLR models that produce location-specific nowcasts and forecasts relative to a model that uses the same regression coefficients $\tilde{\gamma}_{vp}$ in all locations, and thus makes the same prediction for each state. This baseline MLR model assumes
\begin{align*}
\boldsymbol c^{(s)}_{lt}  \mid n^{(s)}_{lt}, \tilde{\boldsymbol \theta}_{t}&\stackrel{indep.}{\sim} \text{Multinomial}\left(n^{(s)}_{lt}, \left(\frac{\tilde{\theta}_{t1}}{1 + \sum_{v = 1}^{V-1} \tilde{\theta}_{tv}}, \dots,\frac{\tilde{\theta}_{t(V-1)}}{1 + \sum_{v = 1}^{V-1} \tilde{\theta}_{tv}}, \frac{1}{1 + \sum_{v = 1}^{V-1} \tilde{\theta}_{tv}}\right)\right), 
\end{align*}
where $\tilde{\theta}_{tv}  = \text{exp}\left\{\tilde{\gamma}_{v1} + \tilde{\gamma}_{v2} t\right\}$ for $v=1,\dots, V - 1$ with uniform priors on $\tilde{\gamma}_{vp}$ for $p = 1, 2$ and $v = 1, \dots, V-1$.

We fit all HMLR models and the baseline model using Markov Chain Monte Carlo as implemented in STAN, which yields samples from the posterior distribution of the unnormalized variant prevalences \citep{gelman2015stan}. 
Each model is fit using one chain with 7000 burn-in iterations and 15000 iterations total.

Having fit all 13 models to each of the 106 datasets corresponding to distinct submission dates denoted by $s$, we obtain nowcasts for the 31 days up to and including the submission date and forecasts for 10 days after the submission date. For each model $m$ fit to the data from submission date $s$, we store the posterior mean variant prevalences $\bar{\boldsymbol \pi}^{(s)}_{mlt\cdot}$ for location $l = 1, \dots, L$, and target date $t$, which refer to the average of $\pi_{ltv} = \theta_{ltv}/(1 + \sum_{v = 1}^{V-1}\theta_{ltv})$ across posterior samples for HMLR models and the average of $\tilde{\pi}_{ltv} =\tilde{\theta}_{tv}/(1 + \sum_{v = 1}^{V-1}\tilde{\theta}_{tv})$ across posterior samples for the MLR model. 
We also store a subset of 100 samples from the posterior distribution of the variant prevalences - $\boldsymbol \pi_{lt\cdot}=(\pi_{lt1},\dots, \pi_{ltV})$ for the HMLR models and $\tilde{\boldsymbol \pi}_{t\cdot} = (\tilde{\pi}_{t1},\dots, \tilde{\pi}_{tV})$ for the MLR model. We denote the stored subsets of variant prevalences for model $m$ fit to data from submission date $s$ at location $l$ and target date $t$ as $\mathcal{P}^{(s)}_{mlt}$. 

We obtain nowcasts and forecasts for variant counts by simulating 100 multinomial random variables per variant proportion drawn from the posterior distribution stored in $\mathcal{P}^{(s)}_{mlt}$ using totals $n^{(s+91)}_{lt}$ that describe the total number of samples observed at each location and time point within 91 days of submission date. 
Stored multinomial draws for model $m$ fit to data from submission date $s$ at location $l$ and target date $t$ are denoted as $\mathcal{C}^{(s)}_{mlt}$. 
For dates and locations for which sequences were observed within 91 days of the submission date that the model was fit to, $n^{(s + 91)}_{lt} > 0$, each element of $\mathcal{C}^{(s)}_{mlt}$ is a draw from a multinomial distribution with total $n^{(s+91)}_{lt}$ and $\mathcal{C}^{(s)}_{mlt}$ contains $|\mathcal{C}^{(s)}_{mlt}|=10,000$ elements. For dates and locations for which no sequences were observed within 91 days of the submission date that the model was fit to, $n^{(s + 91)}_{lt} = 0$, i.e.\ $\mathcal{C}^{(s)}_{mlt}$ is an empty set with $|\mathcal{C}^{(s)}_{mlt}|=0$ elements.

\subsection{Model Evaluation}\label{sec::scores}

Having fit models and constructed nowcasts and forecasts, we are tasked with the challenge of finding a suitable scoring rule for evaluating nowcasts and forecasts. To evaluate the nowcast and forecast distributions and the average nowcasts and forecasts, we follow the lead of the SARS-CoV-2 Variant Nowcast Hub in terms of scoring rules \citep{Variant-nowcast2024}, which specifies that forecast and nowcast distributions are scored using the energy score\citep{gneiting2008assessing, jordan2019evaluating}, whereas the posterior means are scored using the Brier score  \citep{glenn1950verification}.

The energy score is a negatively oriented proper scoring rule \citep{gneiting2008assessing} that scores forecasts based on two components: accuracy, how close nowcasts or forecasts are to the observed data on average, and sharpness, how variable the nowcasts or forecasts are. Given a nowcast or forecast distribution $\mathcal{F}$ containing $|\mathcal{F}|$ samples and an observed value $\boldsymbol z$,  the approximate energy score $ES \left (\mathcal{F},\boldsymbol z \right )$ based on the set of samples $\mathcal{F}$ is defined as
\begin{equation}
 ES \left (\mathcal{F},\boldsymbol z \right ) =
\frac{1}{\left|\mathcal{F}\right|}\sum_{\boldsymbol f \in \mathcal{F}}||\boldsymbol z - \boldsymbol f|| - 
\frac{1}{2\left|\mathcal{F}\right|^2}\sum_{\boldsymbol f \in \mathcal{F}}\sum_{\boldsymbol f' \in \mathcal{F}}||\boldsymbol f - \boldsymbol f'||, 
\label{eq:essum}   
\end{equation}
where $||\boldsymbol f|| =\boldsymbol f'\boldsymbol f$ is the sum of squared elements of $\boldsymbol f$.

The energy score for model $m$, dataset indexed at submission date $s$, location $l$, and target date $s - h$ that is $h$ days out from the submission date $s$, given a collection of samples $\mathcal{C}^{(s)}_{ml(s+h)}$ and given observed variant counts $\boldsymbol c^{(s + 91)}_{l(s+h)}$ is defined as
\begin{equation}
ES_{mlsh} = ES \left (\mathcal{C}^{(s)}_{ml(s+h)}, \boldsymbol c^{\left(s+91\right)}_{l(s+h)} \right ).
\label{es:oneday}
\end{equation}
To describe how we average energy scores within models across locations and submission dates, we introduce additional notation to describe the locations and dates for which non-zero variant counts are observed within $91$ days of the submission date. Let $\mathcal{H}_{ls} = \{h: n^{(s+91)}_{l(s+h)} > 0 \text{ and } -31 \leq h \leq 10\}$ be the set of horizons $h$ between $-31$ and $10$ relative to a submission date $s$ for which non-zero variant counts are observed within $91$ days of the submission date. This is the set of all horizons that are scored for that submission date-location combination.
We define the mean energy score for model $m$, a submission date $s$, and location $l$ across horizons $h$ as $ES_{mls\cdot} = (1/|\mathcal{H}_{ls}|)\sum_{h \in \mathcal{H}_{ls}}ES_{mlsh}$. We define the mean energy score for model $m$ and location $l$ across horizons $h$ and submission dates $s$ as 
\begin{align*}
    ES_{ml\cdot \cdot} = \frac{1}{\sum_{s=1}^{106} |\mathcal{H}_{ls}|}\sum_{s=1}^{106}\sum_{h \in \mathcal{H}_{ls}}ES_{mlsh},
\end{align*}
the mean energy score for model $m$ and submission date $s$ across horizons $h$ and locations $l$ as 
\begin{align*}
    ES_{m\cdot s \cdot} = \frac{1}{\sum_{l=1}^{L} |\mathcal{H}_{ls}|}\sum_{l=1}^{L}\sum_{h \in \mathcal{H}_{ls}}ES_{mlsh},
\end{align*}
 and the overall mean energy score for model $m$ across all locations $l$, submission dates $s$, and horizons $h$ as 
\begin{align*}
    ES_{m\cdot \cdot \cdot} = \frac{1}{\sum_{l=1}^{L} \sum_{s=1}^{106}|\mathcal{H}_{ls}|}\sum_{l=1}^{L}\sum_{s = 1}^{106}\sum_{h \in \mathcal{H}_{ls}}ES_{mlsh}.
\end{align*}

We use Brier scores for categorical multivariate data to score the posterior means. The Brier score for model $m$, dataset indexed at submission date $s$, location $l$, and target date $s - h$ that is $h$ days out from the submission date $s$, given posterior mean nowcast/forecast variant prevalences  $\bar{\boldsymbol \pi}^{(s)}_{ml(s+h)\cdot}$ and observed variant prevalences $\boldsymbol p^{(s + 91)}_{l(s+h)\cdot} = (c^{(s+91)}_{lt1}/n^{(s+91)}_{l(s+h)}, \dots, c^{(s+91)}_{ltV}/n^{(s+91)}_{l(s+h)})$ is defined as 
\begin{equation}
BS_{mlsh} = \frac{1}{2}\sum_{v = 1}^V p^{(s + 91)}_{l(s+h)v} \left(\bar{\pi}^{(s)}_{ml(s+h)v} - 1\right)^2 + \left(1 - p^{(s + 91)}_{l(s+h)v}\right)\left(\bar{\pi}^{(s)}_{ml(s+h)v}\right)^2.
\label{eq:our_Brier}
\end{equation}
This score is based on Brier's original formulation \citep{glenn1950verification}.
We aggregate Brier scores for a single model across horizons $h$, submission dates $s$, and locations $l$ as we aggregated energy scores, with $BS_{mls\cdot} = (1/|\mathcal{H}_{ls}|)\sum_{h \in \mathcal{H}_{ls}}BS_{mlsh}$ and
\begin{align*}
     BS_{ml\cdot \cdot} &= \frac{1}{\sum_{s=1}^{106} |\mathcal{H}_{ls}|}\sum_{s=1}^{106}\sum_{h \in \mathcal{H}_{ls}}BS_{mlsh}, \quad BS_{\cdot s \cdot} = \frac{1}{\sum_{l=1}^{L} |\mathcal{H}_{ls}|}\sum_{l=1}^{L}\sum_{h \in \mathcal{H}_{ls}}BS_{mlsh}, \text{ and } \\
     BS_{m\cdot \cdot \cdot} &= \frac{1}{\sum_{l=1}^{L} \sum_{s=1}^{106}|\mathcal{H}_{ls}|}\sum_{l=1}^{L}\sum_{s = 1}^{106}\sum_{h \in \mathcal{H}_{ls}}BS_{mlsh}.
\end{align*}

We also consider median energy and Brier scores across all locations, submission dates, and horizons for which energy or Brier scores can be computed.

\section{Results}

\subsection{Case Study of One Location and Submission Date}\label{sec::case}

\subsubsection{Model Specifications Impact Shape and Uncertainty}

\begin{figure}[H]
    \centering
    \includegraphics[width=0.9\columnwidth,height=0.8\textheight, keepaspectratio]{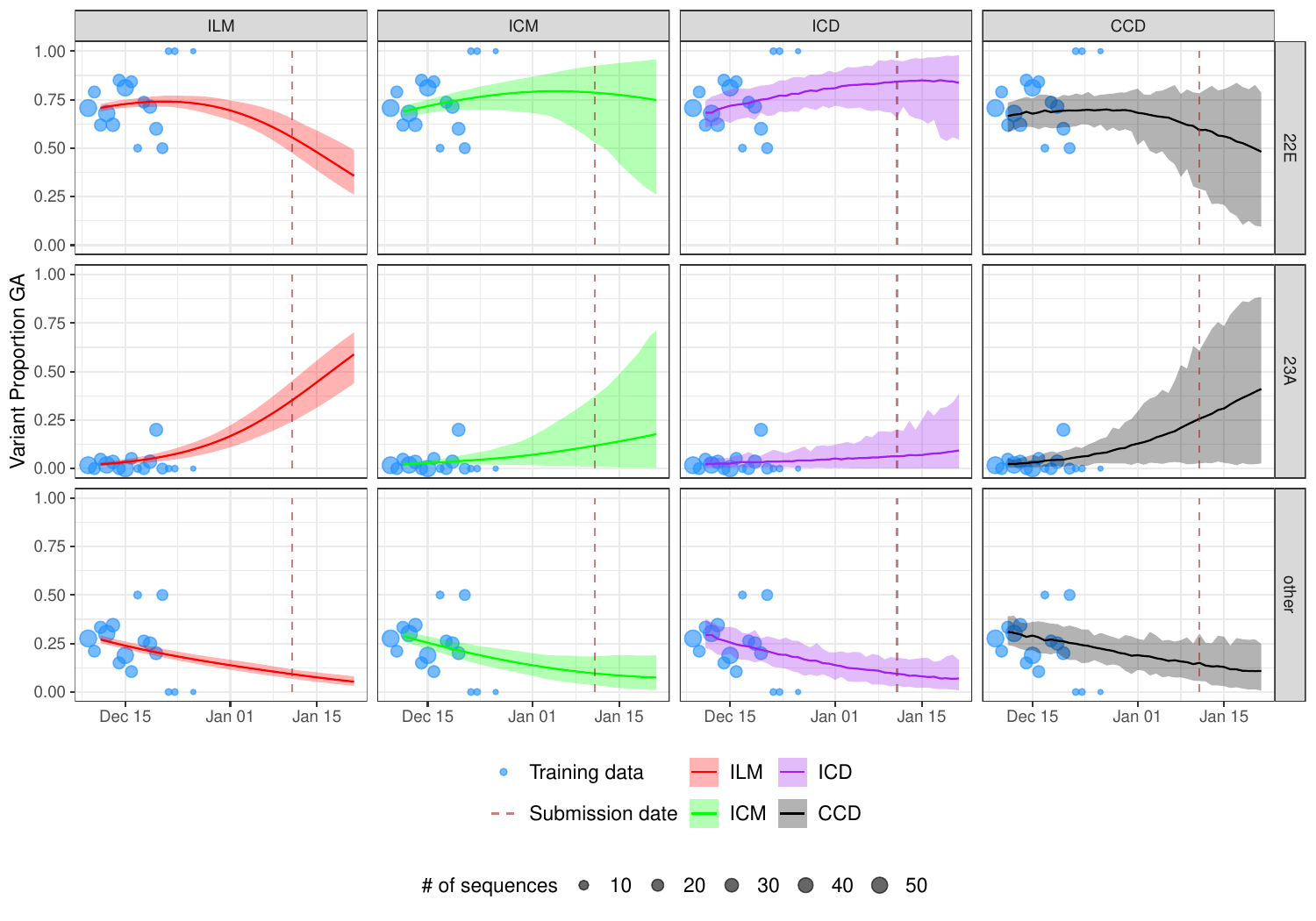}
\caption{Forecasts and nowcasts for a subset of four HMLR models at a single location (Georgia) and submission date (January 11, 2023). For each model, observed variant fractions $p^{(s)}_{ltv} = c^{(s)}_{ltv}/n^{(s)}_{lt}$ observed by the submission date for the nowcast period are shown as blue dots, posterior mean nowcast/forecast variant prevalences denoted as solid lines, $\bar{\boldsymbol \pi}^{(s)}_{mlt}$ are depicted along with approximate 90\% credible intervals for nowcast/forecast variant prevalences, shown as bands, which are obtained by computing the 5\% and 95\% sample quantiles of variant prevalences stored in $\mathcal{P}^{(s)}_{mlt}$. The four selected models are chosen to highlight how the assumed time trend, distribution of variant counts, and regression coefficient covariances structure affect average variant proportion nowcasts and forecasts and their variability. Models are arranged from simplest, a multinomial model with a linear time trend and independent regression coefficients with separate variances (ILM), to most complex, a Dirichlet-multinomial model with a cubic time trend and correlated regression coefficients (CCD).}
    \label{fig:diff_across_models}
\end{figure}

Studying one set of forecasts and nowcasts of variant proportions, made on 2023-01-11 for Georgia, illustrates typical distinctions between models (Figure \ref{fig:diff_across_models}). 
On this date,  the simplest and most complex models (ILM and CCD, respectively) yield substantively similar nowcasts and forecasts in terms of mean prediction, but with large differences in the amount of uncertainty around the mean. Both indicate a steep decline in 22E and a sharp rise in 23A. In contrast, the cubic models that assume independent regression coefficients (ICM and ICD) both indicate a milder decline in 22E and gentler rise in 23A. 
Going from ILM to ICM, and thus from a linear to cubic model, leads to a large difference in mean predictions and uncertainty. These changes are because we are now using a different assumed polynomial over time. 
Changing from ICM to ICD, and thus from a multinomial to a Dirichlet-multinomial model, leads to a reduction in the amount of uncertainty, especially for later horizons and a small change in mean proportions. While the Dirichlet-multinomial model having less uncertainty than the multinomial model may seem counterintuitive, in this case, we find that the estimated over-dispersion parameter is low, with very similar dispersion to a multinomial model, and the Dirchlet-multinomial model has less variance among the sample $\tilde{\boldsymbol \theta}_{lv}$, particularly for the clade $23$A. This suggests that an additional benefit of Dirichlet-multinomial models, beyond the potential to accommodate overdispersed counts, is their ability to produce more stable estimates of the overall mean trend in the presence of extreme values and limited data.        
Moving from ICD to CCD, and thus from an individual to correlated, leads to a rise in uncertainty and a dramatic shift in the mean predictions. The full correlation matrix leads to more uncertainty in predictions by allowing trends for each variant to depend more on the observed behavior of and uncertainty about all variants.        

\subsubsection{Model Specifications Impact Performance}

\begin{figure}[h]
    \centering
\includegraphics[width=0.9\columnwidth,height=0.8\textheight, keepaspectratio]{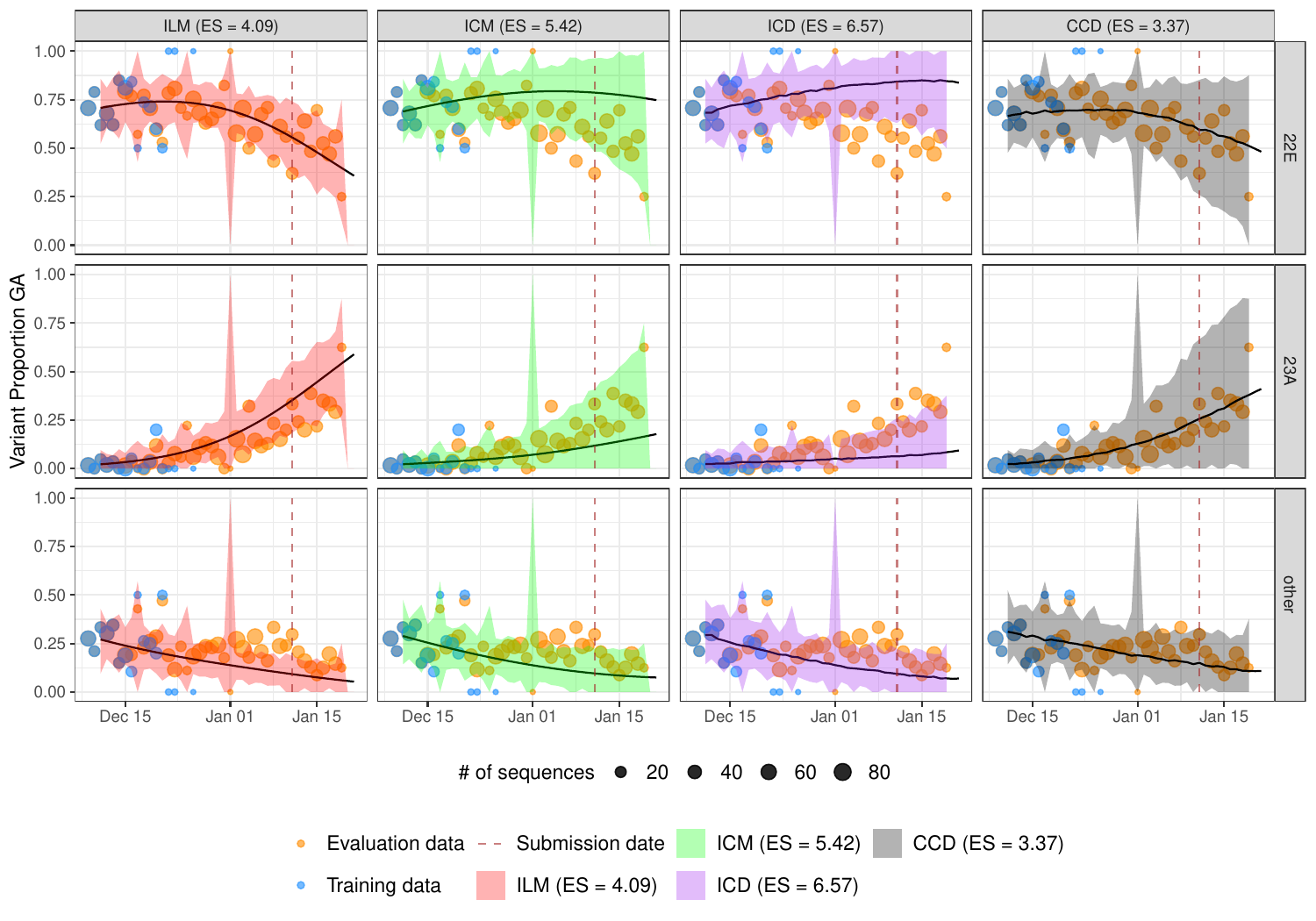}
\caption{Forecasts and nowcasts for Georgia for the submission date of 2023-01-11.  Solid lines denote posterior mean variant prevalences, and bands denote approximate 90\% prediction intervals for nowcast/forecast variant proportions, which are obtained by computing the 5\% and 95\% sample quantiles of posterior predictive variant fractions stored in $\mathcal{C}^{(s)}_{mlt}$. Mean energy scores for each model are displayed on the label energy scores, here averaged across horizons $ES_{mls\cdot}$ for this location and submission date. Blue dots are variant fractions observed by the submission date, and yellow dots are the variant fractions observed 91 days after the submission date.}
    \label{fig:GA_PI}
\end{figure}

All four models produce prediction intervals of varying widths that reflect not only the variability of nowcast and forecast variant prevalences (Figure~\ref{fig:diff_across_models}) but also the variability in the number of sequences observed per day within $91$ days of the submission date, $n^{(s+91)}_{lt}$. We observe some wide intervals, e.g. on January 1, that range from 0 to 1 when the corresponding sequence count $n^{(s+91)}_{lt}$ is especially low, thereby yielding central intervals of simulated counts that include all and none of the counts belonging to the one clade.

Visually, the simplest and most complex models, ILM and CCD, produce better nowcasts and forecasts that correctly identify the eventually observed decline in 22E and rise in 23A. Better coverage of the prediction intervals by the CCD model, compared to the ILM model, suggests that the CCD model in particular provides the best performance out of the four models shown in Figure~\ref{fig:GA_PI}. The ICD model provides the poorest nowcasts and forecasts with a mean forecast that misses the trends and narrow predictive uncertainty that fails to cover many observations.

The visual assessment of model performance is consistent with the energy scores which are computed jointly across variants and averaged across horizons $ES_{mls\cdot}$ (see Section~\ref{sec::scores}). CCD has the best (lowest) score, followed by ILM, which provides the same expected forecasts and nowcasts but has narrower prediction intervals with poorer coverage. Both ICM, which fails to capture the variant proportion trends correctly, and ICD, which both fails to capture the variant proportion trends correctly and underestimates variability, have much higher mean energy scores. Overall, the results suggest that, for this specific location-date combination, allowing regression coefficients to be correlated while adding complexity by using a higher degree polynomial trend (features of the CCD model) helps regularize coefficient estimates that are otherwise challenging to estimate well from the available data.

\subsection{HMLR Outperforms MLR Baseline Overall}

Overall performance of the HMLR models and the baseline MLR model is summarized in Table~\ref{tab:overall_scores} across all locations, submission dates, and horizons as measured by mean and median energy score, and mean and median Brier score. 
We consider both mean and median scores across locations, submission dates, and horizons because we observe that the distributions of energy scores across locations, submission dates, and horizons are right-skewed, with all models having extremely poor performance for some locations, submission dates, and horizons. For this reason, the mean energy score is more sensitive to performance for the most challenging locations, submission dates, and horizons to produce nowcasts and forecasts for. The Brier scores do not show this right skew with more similar values for mean and median scores. We consider both energy and Brier scores because they allow us to separately characterize the quality of our nowcast and forecast distributions overall - including their variability - and the quality of our posterior mean nowcasts and forecasts as point estimates.

\begin{table}[h]
\centering 
\caption{Mean and Median Energy and Brier Scores.}
    \begin{tabular}{|l|l|l|l|l|}\hline
    \multirow{2}{*}{Model} & \multicolumn{2}{c|}{Energy Score} & \multicolumn{2}{c|}{Brier Score} \\
         & Mean & Median & Mean & Median  \\
        \hline
        SLM & $2.275$ & $1.203$ & $0.263$ & $\boldsymbol{0.272}$\\
        SCM & $3.046$ & $1.314$  & $0.275$ & $0.290$\\
        \hline
        ILM & $2.284$ & $1.203$ & $0.263$ & $0.273$\\
        ICM & $2.104$ & $1.222$ & $0.266$ & $0.277$\\
        \hline
        CLM & $2.277$ & $\boldsymbol{1.199}$ & $0.262$ & $\boldsymbol{0.272}$\\
        CCM & $2.056$ & $1.205$ & $0.265$ & $0.275$\\
        \hline
        SLD & $2.194$ & $1.210$ & $\boldsymbol{0.261}$ & $\boldsymbol{0.272}$\\
        SCD & $2.070$ & $1.233$ & $0.266$ & $0.278$\\
        \hline
        ILD & $2.203$ & $1.210$ & $0.262$ & $\boldsymbol{0.272}$\\
        ICD & $2.050$ & $1.226$ & $0.265$ & $0.276$\\
        \hline
        CLD & $2.195$ & $1.210$ & $0.262$ & $\boldsymbol{0.272}$\\
        CCD & $\boldsymbol{1.998}$ & $1.205$ & $0.264$ & $0.274$\\
        \hline
        Baseline & $2.739$ & $1.2567$ & $0.268$ & $0.282$ \\ \hline
    \end{tabular}
    \label{tab:overall_scores}
    \caption{Means and medians are taken over all submission dates, locations and horizons, i.e $ES_{m\cdot \cdot \cdot}$. Best scores are \textbf{bolded}.}
\end{table}
% Analysis by model

 Table~\ref{tab:overall_scores} clearly demonstrates the value of HMLR models relative to an MLR baseline model for this data and problem. All but one HMLR model - the multinomial model that assumes a cubic trend and that regression coefficients are independent with a common, shared variance (SCM) - outperform the baseline MLR model according to all four metrics, although the differences are sometimes small.

 Which specific HMLR  model is best, and which model specifications tend to improve performance, depends on how performance is measured. The mean energy score, which takes the dispersion of the nowcast and forecast distributions into account and is sensitive to high (poor) scores at the most challenging locations, submission dates, and horizons to produce forecasts for, tends to prefer more dispersed Dirichlet-multinomial models for variant counts over multinomial models and favors models with more complex cubic time trends versus linear time trends. The top-performing model, according to mean energy score, is the most complex one; the Dirichlet-multinomial model with a cubic time trend that allows regression coefficients to be correlated (CCD). Median energy score is less sensitive to extreme scores, which can occur at the most challenging locations, submission dates, and horizons. The mean and median Brier scores, which do not take dispersion of nowcasts and forecasts into account, tend to favor simpler models for variant counts with linear time trends that assume independence and shared variances across regression coefficients. Specifically, the median energy score favors multinomial models (SLM and ILM), the mean Brier score favors the Dirichlet-multinomial models (SLD), and the median Brier score ranks most HMLR models very similarly and favors the models with linear time trends (SLM, CLM, SLD, ILD, and CLD).

 In what follows, we focus on performance as measured by mean energy score, as opposed to median energy score or mean or median Brier score. This reflects our interest in obtaining accurate nowcast and forecast distributions as opposed to just nowcast and forecast point estimates and our interest in improving nowcasts and forecasts at the locations, submission dates, and horizons that are most challenging to produce nowcasts and forecasts for.

\subsection{HMLR Outperforms Baseline Across Time, Particularly in 2022}

To better understand when different HMLR model specifications are especially advantageous, we consider mean energy scores at distinct submission dates averaged across locations and horizons. We compare the performance of all 13 models across all submission dates in Figure~\ref{fig:scores_plot1}. Letting $m = 0$ refer to the baseline model, we summarize model performance in the following figures using the log relative energy scores compared to the baseline model, $\text{log}(ES_{m\cdot s \cdot}/ES_{0\cdot s \cdot})$, where $ES_{m\cdot s \cdot}$ is the mean energy score for model $m$ and submission date $s$ across all locations and horizons as defined in Section~\ref{sec::scores}.

\begin{figure}[h]
\centering
\includegraphics[width=0.7\columnwidth,height=\textheight, keepaspectratio]{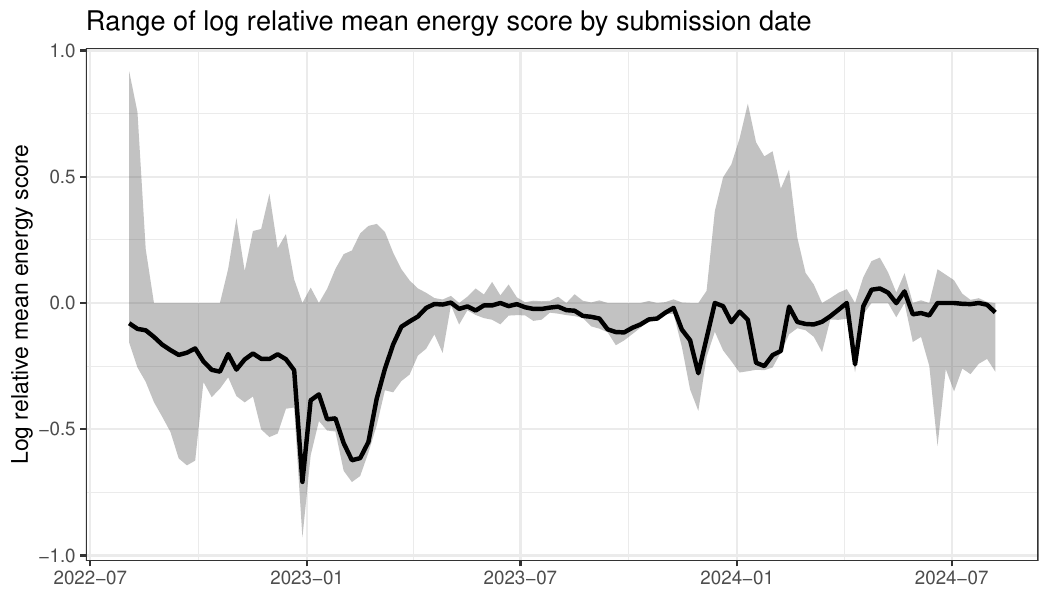}
\caption{Median and range of log relative energy score of HMLR models versus the MLR baseline by submission date. Lower scores are better. The median is the dark line, and the shaded region bounds the highest and lowest relative mean energy score. If a model's score is below 0, that model is outperforming the baseline.}
\label{fig:scores_plot1}
\end{figure}

 Figure~\ref{fig:scores_plot1} shows the greatest benefits from HMLR models relative to the MLR baseline during the second half of 2022 and first half of 2023, little variation in model performance across all 13 models during the second half of 2023, and some benefits from HMLR models relative to the MLR baseline in 2024.

\begin{figure}[H]
\centering
\includegraphics[width=\columnwidth,height=0.5\textheight, keepaspectratio]{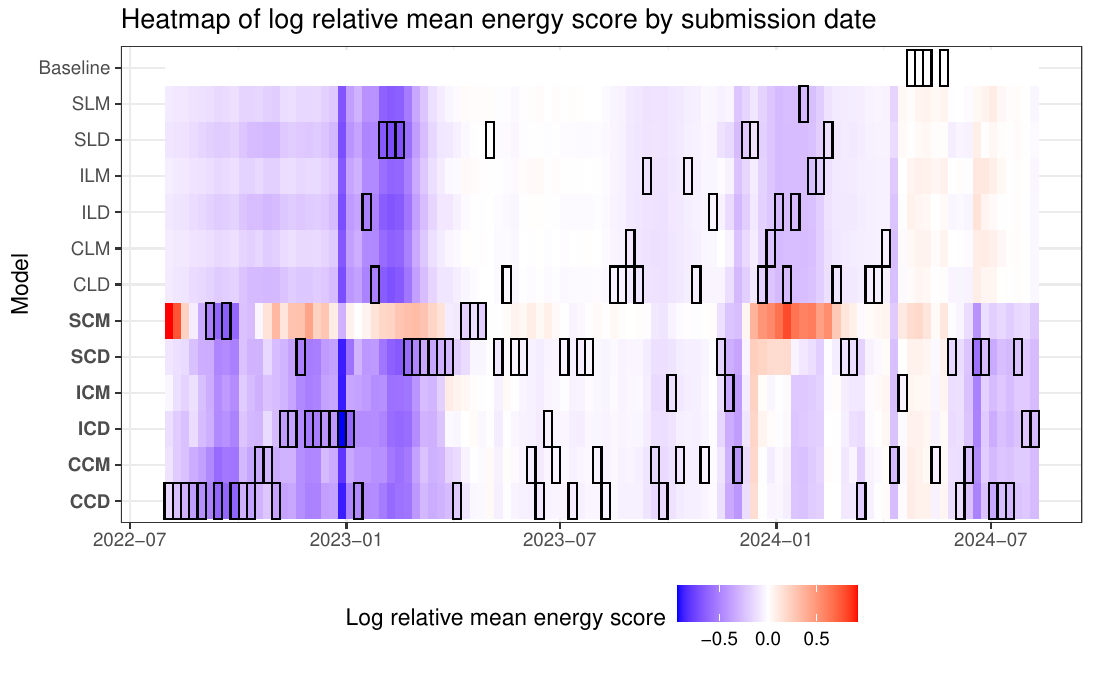}
\caption{A heatmap of log relative mean energy scores by submission date with the best-scoring model for each week outlined in black. Scores less than 0 (blue) indicate the model performed better than the baseline in the given week. Scores greater than 0 (red) indicate the model performed worse than the baseline.}
\label{fig:scores_plot2}
\end{figure}

The best model for each submission date and HMLR model-specific log relative energy scores relative to the MLR baseline are summarized in Figure~\ref{fig:scores_plot2}. Every model was the best performing model for at least one submission date and HMLR models performed best most often, on all but four submission dates. With the exception of the multinomial model that assumes a cubic trend and assumes that regression coefficients are independent with a common, shared variance (SCM), all other HMLR models tend to outperform the MLR baseline most days, especially in the second half of 2022 and the first half of 2023. The biggest systematic difference in performance is between the linear and cubic HMLR models. The cubic models tend to perform better than their linear counterparts and, unlike their linear counterparts, tend to outperform the MLR baseline during the second half of 2024.

\subsection{HMLR Outperforms MLR Baseline More in States with More Data}

\begin{figure}[H]
\begin{center}
\includegraphics[width=\columnwidth,height=\textheight, keepaspectratio]{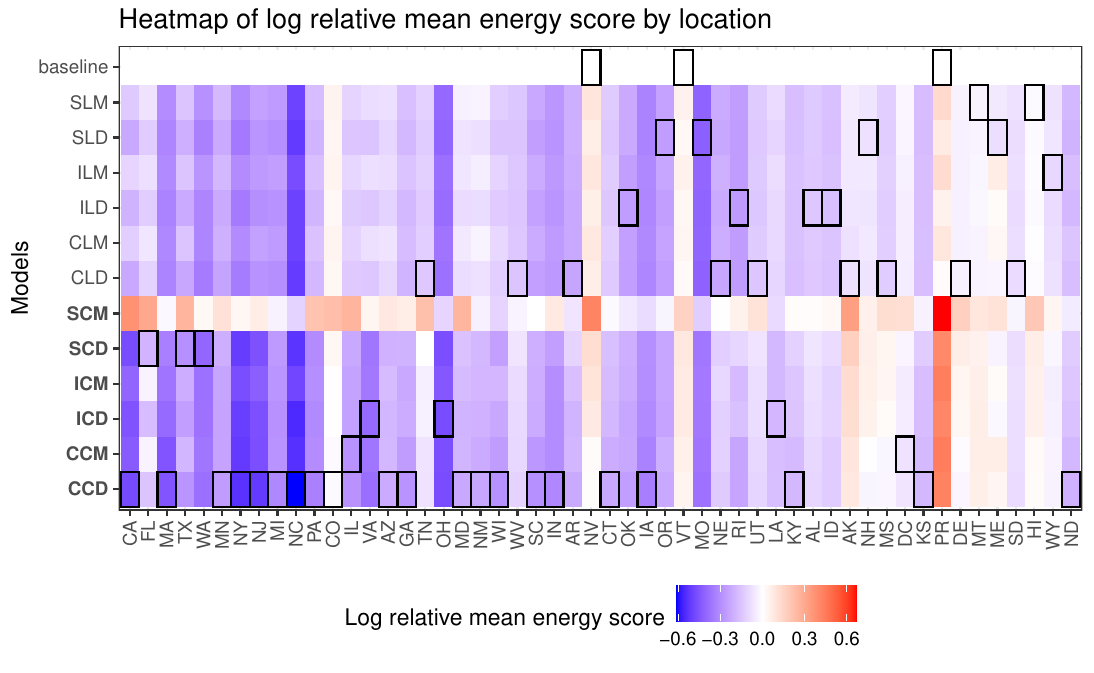}
\caption{A heatmap showing the relative energy scores across locations, averaged over submission date and horizon best-scoring model for each week outlined in black. Locations are ordered in descending order by the amount of data. Scores less than 0 (blue) indicate the model performed better than the baseline in the given week. Scores greater than 0 (red) indicate the model performed worse than the baseline.}
\label{plot::Model_scores_across_states}
\end{center}
\end{figure}

We also consider mean energy scores at distinct locations, averaged across submission dates and horizons.
Figure~\ref{plot::Model_scores_across_states} shows log relative mean energy scores for each HMLR model compared to the  MLR baseline across 52 states, with states ordered according to the amount of data available, from most to least. Log relative mean energy scores are defined as $\text{log}(ES_{ml \cdot \cdot}/ES_{0 l \cdot \cdot})$, using the model and location-specific mean energy scores $ES_{m l \cdot \cdot}$ described in Section~\ref{sec::scores}.

From Figure~\ref{plot::Model_scores_across_states}, most of the HMLR models outperform the MLR baseline in terms of mean energy score across most of the locations, especially at locations with more data. The baseline is only the best model in just three of 52 locations.
Trends in location-specific relative HMLR model performance shown in  Figure~\ref{plot::Model_scores_across_states} are similar to the trends in overall HMLR model performance previously shown in Table~\ref{tab:overall_scores}. Dirichlet-multinomial HMLR models tend to outperform their multinomial HMLR counterparts, with Dirichlet-multinomial HMLR models performing best at 44 of 52 locations. The HMLR models that performs best and worst across the most locations is the same as the best and worst performing model according to mean energy score overall; the Dirichlet-multinomial model with a cubic time trend that allows regression coefficients to be correlated (CCD) is the top performing model and the the multinomial model that assumes a cubic trend and that regression coefficients are independent with a common, shared variance (SCM) performs poorly across all states.

The trend of HMLR models outperforming the MLR baseline less at locations with less data reflects both the HMLR models' tendancy to shrink towards an across-location average model that is similar to the MLR baseline at locations where very little data is available, and the difficulty of comparing model performance when very little validation data becomes available even after 91 days, i.e., when specimen counts $n^{(s + 91)}_{lt}$ are low. The latter challenge is reflected in Figure~\ref{fig:GA_PI}; when $n^{(s + 91)}_{lt}$ is small, the nowcast and forecast distributions can be very variable, even for models that produce precise estimated variant prevalences.
Similarly, the trend of HMLR models with a cubic time trend only outperforming their linear time trend counterparts when relatively more data is available reflects the need to observe more data to fit more complex HMLR models and the challenges of comparing models when limited data is available.

\section{Discussion}
\label{sec::Disc}
The state-level HMLR models, with one exception, outperform the country-level MLR baseline in all the metrics tested, with the largest improvement coming in terms of mean energy score. This suggests that state-level HMLR models may be more applicable for probabilistic modeling of SARS-CoV-2 than a country-level MLR model. Interestingly, the largest improvements of state-level HMLR models compared the country-level MLR model were seen in states with the most data compared to the country level HMLR models that found the largest improvement in states with the least data \citep {Abousamra2024, Susswein2023}.

Overall, Cubic Dirichlet-multinomial models were found to perform the best in terms of mean energy score, whereas simpler linear models performed best in terms of Brier score. These results suggest that HMLR techniques may be useful for modeling SARS-CoV-2 variant prevalences both in situations both in situations where many locations have a lot of data as well as situations where there are large differences across locations. However, the particular type of HMLR model that is best will depend on what is the particular scientific question considered; in particular, what scoring metric is being used.    

There are a few caveats to this analysis. The predictions are also made on historical data; the quantity of sequence data generated in 2022 and 2023 is much higher than that in more recent years, so this may limit how much we can generalize the results. The sequences used as data were also only the cases that were tested at a lab, so there is no guarantee that they are an unbiased random sample of the true population. As explained in Section~\ref{sec::dataset}, the retrospective datasets were constructed by assigning sequences to clades using the probabilistic clade assignment model and corresponding tree that was in operation on November 22, 2024 \citep{Roemer2024Tree}, regardless of submission date. If we had been constructing the same datasets in real time, we would have assigned sequences to clades using the probabilistic clade assignment model that was available at the submission date associated with each dataset. Thus, the clades modeled for certain weeks in the retrospective datasets may not be the same clades that would have been modeled if the Variant Nowcast Hub were running in 2022. After this analysis was started, the tool CladeTime was created to retrieve historical versions of Nextstrain probabilistic clade models and reclassify sequences using those models \citep{CladeTime}. However, this data is only available from September 2024, so future work could involve replicating this analysis with times starting in September 2024 using CladeTime.

There is much more testing required to determine if there is a specific type of HMLR model that is generally best for modeling SARS-CoV-2 and the benefits of HMLR models over MLR models in other settings involving genomic data.
Possible future lines of inquiry include the development of additional models that allow for more flexible time trends, e.g. via use of splines to define time trends, incorporation of additional covariates, use of location-specific regression coefficient variances, and relaxation of the independence assumption across locations and predictors by using spatially correlated models for regression coefficients that allow for more flexible information sharing across locations. Approaches that use machine learning instead of regression-based approaches could also be explored.  Future work may also consider the use of HMLR versus MLR models for other infectious diseases.

\subsection{Code Availability}
Code and data for generating the results in this paper can be found on GitHub at \url{https://github.com/reichlab/variant-nowcast-model-dev-retro}.

\section*{Acknowledgments}

This work has been supported by the National Institutes of General Medical Sciences (R35GM119582) and the US Centers for Disease Control and Prevention (U01IP001122). The content is solely the responsibility of the authors and does not necessarily represent the official views of NIGMS, the National Institutes of Health, or CDC.
NGR discloses consulting income from Google Research unrelated to this manuscript.
We would like to acknowledge Kaitlyn Johnson for her thoughtful feedback on this work, as well as the Nextstrain team for curating the data that makes this work possible.
\bibliography{references}
%\printbibliography
\end{document}